\documentclass{article}

\usepackage[preprint]{neurips_2025}

\usepackage[utf8]{inputenc} 
\usepackage[T1]{fontenc}    
\usepackage{hyperref}       
\usepackage{url}            
\usepackage{booktabs}       
\usepackage{amsfonts}       
\usepackage{nicefrac}       
\usepackage{microtype}      
\usepackage{xcolor}         

\usepackage{graphicx}
\usepackage{subcaption}
\usepackage{amsmath}
\usepackage{multirow}
\usepackage{multibib}
\usepackage{wrapfig}

\newcites{app}{Appendix references}

\newcommand{\arcond}{X^{t-2:t-1}}
\newcommand{\vardist}{q\left(Z^t \middle| X^{t-2:t-1}, X^{t} \right)}
\newcommand{\priordist}{p\left(Z^t \middle| \arcond \right)}

\newcommand{\varloss}{\mathcal{L}_\mathrm{Var}}
\newcommand{\crpsloss}{\mathcal{L}_\mathrm{CRPS}}

\title{Graph-based Neural Space Weather Forecasting}

\author{%
  \textbf{Daniel Holmberg}$^{1,2}$\thanks{Corresponding author: \texttt{dholmberg@ucsb.edu}}\quad
  \textbf{Ivan Zaitsev}$^{1}$\quad
  \textbf{Markku Alho}$^{1}$\quad
  \textbf{Ioanna Bouri}$^{1}$\quad
  \textbf{Fanni Franssila}$^{1}$\quad\\
  \textbf{Haewon Jeong}$^{2}$\quad
  \textbf{Minna Palmroth}$^{1,3}$\quad
  \textbf{Teemu Roos}$^{1}$\\\\
  $^{1}$University of Helsinki\quad$^{2}$UC Santa Barbara\quad$^{3}$FMI Space and Earth Observation Centre
}

\begin{document}

\maketitle

\begin{abstract}
Accurate space weather forecasting is crucial for protecting our increasingly digital infrastructure. Hybrid-Vlasov models, like Vlasiator, offer physical realism beyond that of current operational systems, but are too computationally expensive for real-time use. We introduce a graph-based neural emulator trained on Vlasiator data to autoregressively predict near-Earth space conditions driven by an upstream solar wind. We show how to achieve both fast deterministic forecasts and, by using a generative model, produce ensembles to capture forecast uncertainty. This work demonstrates that machine learning offers a way to add uncertainty quantification capability to existing space weather prediction systems, and make hybrid-Vlasov simulation tractable for operational use.
\end{abstract}

\section{Introduction}

Space weather describes conditions in near-Earth space driven by the solar wind and the internal dynamics of Earth's magnetosphere. These conditions threaten modern infrastructure by creating geomagnetically induced currents that disrupt power networks~\cite{bolduc2002gic, dimmock2019gic}, adversely affecting satellite operations~\cite{gubby2002space, zhang2022thermospheric}, and causing failures in electromagnetic communication~\cite{baker2004effects}. Current operational forecasting relies on driving global magnetohydrodynamic (MHD) models like BATS-R-US~\cite{glocer2013coupledbatsrus} with real-time solar wind data from satellites orbiting the L1 Lagrange point, such as ACE~\cite{stone1998advanced}. However, this paradigm faces several challenges. First, MHD models approximate plasma as a fluid, omitting ion-kinetic processes that are only captured by higher-fidelity but computationally prohibitive simulations like hybrid-Vlasov models. Second, most operational forecasts are deterministic single-point predictions, which lack crucial uncertainty information. To address this, the need for ensemble forecasting has been highlighted~\cite{murray2018importance}, with existing approaches perturbing the solar wind inputs for physics-based models~\cite{morley2018perturbed}, or using machine learning for post-processing deterministic outputs~\cite{camporeale2019generation}.

To tackle these challenges, we take inspiration from recent breakthroughs in machine learning for atmospheric weather forecasting~\cite{keisler2022forecasting, bi2023accurate, lam2023graphcast, oskarsson2024probabilistic, price2025probabilistic, larsson2025diffusion}, and adapt graph-based limited-area modeling~\cite{oskarsson2024probabilistic} to the domain of space weather. We tailor established graph neural network (GNN) architectures~\cite{keisler2022forecasting, lam2023graphcast, oskarsson2024probabilistic} to a magnetospheric simulation grid in the noon-midnight meridional plane, excluding the circular inner boundary close to Earth, and incorporate the upstream solar wind as boundary condition. By doing so, we can produce high-fidelity forecasts of the magnetosphere's evolution more than 100 times faster on 1 GPU than the original simulation on 50 CPUs. We show that the framework is capable of both deterministic and probabilistic forecasting, with the latter using a latent-variable approach to generate ensembles that provide uncertainty information currently missing from most space weather models.

\begin{figure}[h]
    \centering
    \includegraphics[width=\textwidth]{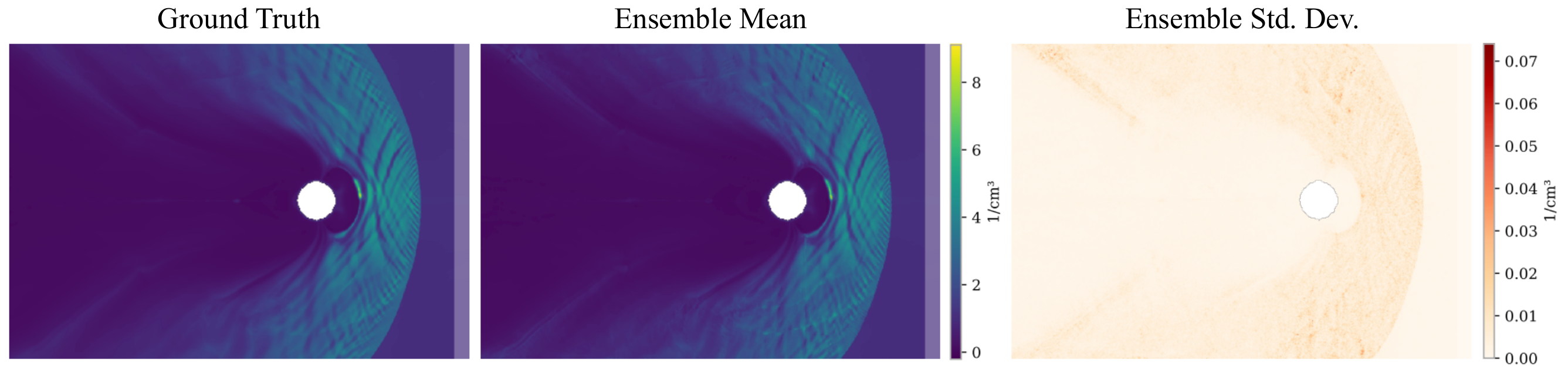}
    \caption{Example forecast of particle density 10 steps ahead, showing the faded inflow boundary.}
    \label{fig:example_forecast}
\end{figure}

\section{Hybrid-Vlasov dataset}

The data for this study were generated using Vlasiator~\cite{vonalfthan2014vlasiator}, which performs global simulations of the solar wind's interaction with the Earth's magnetosphere in the hybrid-Vlasov formalism. Ions are treated as a velocity distribution function, $f$, that depends on position $\mathbf{x}$, velocity $\mathbf{v}$, and time $t$. Its evolution is dictated by the Vlasov equation, which describes how $f$ changes due to the electric field $\mathbf{E}$ and magnetic field $\mathbf{B}$:
\begin{equation}
\frac{\partial f}{ \partial t} + {\bf v} \cdot \frac{\partial f}{\partial {\bf x} } + \left( {\bf E} + {\bf v} \times {\bf B} \right) \cdot \frac{\partial f}{\partial {\bf v } } = 0.
\end{equation}
Electrons are modeled as a massless, charge-neutralizing fluid where the number density $n$ is equal for both ions and electrons ($n_i \simeq n_e \simeq n$). The electromagnetic fields are evaluated by solving the Maxwell-Darwin system:
$
\nabla \times \mathbf{E} = -\partial_t \mathbf{B},\ \nabla \times \mathbf{B} = \mu_0 \mathbf{J},\ \nabla \cdot \mathbf{B} = 0.
$
This system is closed by relating the fields to the moments of the ion distribution function through a generalized Ohm's law that includes the Hall term, which involves the current density $\mathbf{J}$ and the elementary charge $e$:
\begin{equation}
{\bf E} + {\bf v} \times {\bf B} = \frac{ {\bf J} \times {\bf B} }{n e}.
\end{equation}

The run was performed in a 2D-3V configuration (two spatial and three velocity dimensions) using a timestep $\Delta t = 1\,\mathrm{s}$ and spatial resolution of $600\,\mathrm{km}$. The simulation domain covers the noon-midnight meridional plane (the $x$-$z$ plane in Geocentric Solar Ecliptic, or GSE, coordinates), extending from $-60\,R_E$ to $+30\,R_E$ along the $x$-axis and $\pm 30\,R_E$ along the $z$-axis. The solar wind is injected at the $+x$ boundary as a Maxwellian distribution, parametrized by an ion density of $\rho=1/\mathrm{cm}^{3}$, wind velocity of $\mathbf{v}=(-750, 0, 0)\,\mathrm{km/s}$, and a plasma temperature of $T=0.5\,\mathrm{MK}$. The interplanetary magnetic field is directed southward with $\mathbf{B}=(0, 0, -5)\,\mathrm{nT}$. These conditions result in an Alfvén Mach number of $M_A = 6.9$. Plasma can exit through the other edges of the domain, where Neumann boundary conditions are applied. The inner boundary is represented as a perfect conductor located at a radius of 3.7 $R_E$ from the Earth's center. All vector quantities are given in GSE coordinates.

\section{Method}

\paragraph{Problem formulation}
We formulate the problem of space weather forecasting as mapping a set of initial magnetospheric states to a sequence of future states. The input consists of two consecutive states, $X^{-1:0} = (X^{-1}, X^{0})$, to capture first-order dynamics~\cite{lam2023graphcast}, with the goal of predicting the future trajectory $X^{1:T} = (X^{1}, \dots, X^{T})$. Each magnetospheric state $X^t \in \mathbb{R}^{N \times d_x}$ is a high-dimensional tensor representing $d_x$ physical variables present in the hybrid-Vlasov simulation across $N$ grid locations in near-Earth space. The complete feature set is listed in Appendix~\ref{app:data}. Deterministic models tackle the problem by producing a single point, typically mean, estimate of the trajectory $X^{1:T}$, while probabilistic approaches aim to model the full conditional distribution $p(X^{1:T} | X^{-1:0})$.

\paragraph{Graph-based neural forecasting}
Using machine learning, the deterministic forecasting problem can be solved with an autoregressive model applied iteratively to produce a forecast. In the probabilistic case, we sample from the model's output distribution and repeat the process to generate an ensemble of trajectories. For both cases, we use a GNN based on an \emph{encode-process-decode} architecture~\cite{sanchez2020learning} where: 1) grid inputs are encoded onto the mesh representation; 2) a number of GNN layers process this latent representation; 3) the processed data is mapped back onto the original grid to produce the final prediction. Here the GNN predicts the next step as a residual update to the most recent input state, making it an easier learning task compared to predicting the next state directly. For the model to better handle the open boundary on Earth's dayside, we apply boundary forcing as an additional input in the region from $x=27\,R_E$ to the domain edge at $x=30\,R_E$, shown as the faded areas in Fig.~\ref{fig:example_forecast}. A static binary mask indicates which grid nodes to force by replacing with ground truth boundary data after each prediction. In an operational scenario this information would come from conditioning on L1 observations~\cite{glocer2013coupledbatsrus, honkonen2022gumics5}, or from a heliospheric host model~\cite{maharana2024employing}.

\paragraph{Mesh variations}
We consider three mesh architectures for the GNN processor: a \emph{simple} mesh coarser than the simulation domain~\cite{keisler2022forecasting}; a \emph{multiscale} mesh~\cite{lam2023graphcast}; and a \emph{hierarchical} mesh~\cite{oskarsson2024probabilistic}. The multiscale and hierarchical meshes are constructed by recursively creating a sequence of graphs, $\mathcal{G}_L, \dots, \mathcal{G}_1$, at varying resolutions. The multiscale mesh, $\mathcal{G}_{MS}$, uses the nodes from the finest graph, $\mathcal{V}_1$, but connects them with edges from all levels, $\mathcal{E}_{L} \cup \dots \cup \mathcal{E}_{1}$, allowing a single GNN layer to process information across all spatial scales simultaneously. In contrast, the hierarchical approach maintains $L$ distinct graph levels, $\mathcal{G}_1, \dots, \mathcal{G}_L$, where the number of nodes decreases at each level. Information is passed between adjacent levels using dedicated inter-level graphs, $\mathcal{G}_{l,l+1}$ up and $\mathcal{G}_{l+1,l}$ down, to model different spatial scales separately. The mappings between grid and mesh nodes occur through bipartite grid-to-mesh $\mathcal{G}_{G2M}$ and mesh-to-grid $\mathcal{G}_{M2G}$ graphs.

\paragraph{Deterministic model}

We use the deterministic graph-based forecasting model \emph{Graph-FM}~\cite{oskarsson2024probabilistic} to generate point estimate forecasts through the autoregressive mapping $\hat{X}^{t} = f(X^{t-2:t-1})$. We test this model with each of the three mesh architectures, adapting its processor accordingly. When using the hierarchical mesh, a processing step is a complete sweep through the hierarchy. A sweep sequentially applies GNNs to propagate information up from the lowest level $\mathcal{G}_1$ to the highest $\mathcal{G}_L$ and then back down. All upward updates are facilitated by \emph{propagation networks}~\cite{oskarsson2024probabilistic}, while the remaining updates use \emph{interaction networks}~\cite{battaglia2016interactionnet}, with all layers mapping to a latent dimensionality $d_z$. This design leverages the inductive bias of interaction networks to retain information, while propagation networks are more effective at forwarding new information through the graph. We train the deterministic models by minimizing a weighted mean square error (MSE) loss.

\paragraph{Probabilistic model} For probabilistic space weather forecasting we employ the graph-based ensemble forecasting model, \emph{Graph-EFM}~\cite{oskarsson2024probabilistic}. To model the distribution $p(X^{t}|\arcond)$ for a single time step, Graph-EFM uses a latent random variable $Z^t$. This variable acts as a low-dimensional representation of the stochastic elements of the weather system that are not captured by the input states. By conditioning the prediction on this latent variable, the model can efficiently sample different, spatially coherent outcomes. The relationship is defined by the integral:
\begin{equation}
p(X^{t}|\arcond) = \int p(X^{t}|Z^{t}, \arcond) p(Z^{t}|\arcond) dZ^{t}
\end{equation}
where the term $p(Z^{t}|\arcond)$ is the latent map and the term $p(X^{t}|Z^{t}, \arcond)$ is the predictor. The latent map is a probabilistic mapping that takes the previous two weather states as input and produces the parameters for a distribution over the latent variable $Z^t$. Specifically, Graph-EFM uses GNNs to map the inputs to the mean of an isotropic Gaussian distribution, effectively encoding the state-dependent uncertainty into the latent space. The variance is kept fixed. This latent distribution is defined over the  nodes $\mathcal{V}_{L}$ in the top, coarsest level of a given mesh graph, as follows:
\begin{equation}
p(Z^{t}|\arcond) = \prod_{v \in \mathcal{V}_{L}} \mathcal{N}(Z_{v}^{t} | \mu_{Z}(\arcond)_{v}, I)
\end{equation}
 ensuring that the stochasticity is introduced at a low-dimensional, spatially-aware level. The predictor is a deterministic mapping that takes a specific sample of the latent variable $Z^t$, along with the previous weather states, to produce the next weather state $\hat{X}^t$. A sampled value of $Z^t$ is injected at the top level of the graph hierarchy and its influence propagates down through the levels to produce a full, high-resolution, and spatially coherent forecast. The predictor is a deterministic mapping, $g$, that produces the next state $\hat{X}^t$ by adding a predicted residual update, $\tilde{g}$, to the previous state $X^{t-1}$:
\begin{equation}
\hat{X}^{t} = g(Z^{t}, \arcond) = X^{t-1} + \tilde{g}(Z^{t}, \arcond).
\end{equation}
By first sampling a $Z^t$ from the latent map and then passing it through the predictor, the model generates one possible future weather state. Repeating this process creates an arbitrarily large ensemble of forecasts. Drawing from the structure of a conditional Variational Autoencoder (VAE)~\cite{kingma2014vae, sohn2015cvae}, we train the Graph-EFM by optimizing a variational objective equivalent to the negative Evidence Lower Bound (ELBO), which combines a Kullback-Leibler (KL) divergence regularizer with a reconstruction loss. Subsequently, we fine-tune the model by adding a Continuous Ranked Probability Score (CRPS) loss~\cite{gneiting2007strictly, rasp2018neural, pacchiardi2024probabilistic} to the objective, for calibration. Further details on the models and the loss functions are available in Appendix~\ref{app:model}.

\section{Experiments}

To evaluate our models, the Vlasiator simulation is causally split into training, validation, and test sets with durations of 10 minutes, 1 minute, and 1 minute, respectively. This ensures that the evaluation tests for meaningful generalization into the future. Because Vlasiator produces such highly resolved simulations of global near-Earth space, we use a dataset with only a short temporal extent for this proof-of-concept. We train both the deterministic Graph-FM and the probabilistic Graph-EFM using the simple, multiscale, and hierarchical mesh architectures. All models are configured with a latent dimension of $d_z = 64$. For the simple and multiscale graphs, the processor consists of 4 processing layers. To ensure a fair comparison, the hierarchical Graph-FM model uses 2 processing layers, as its sweep mechanism effectively doubles the number of updates per layer. Graph-EFM is set to generate an ensemble size of 5. On a single GPU, the deterministic models predict the next step approximately 500 times faster than the original simulation running on 50 CPUs, while our probabilistic models are roughly 80 times faster. For this speed comparison one should note that the ML models are trained on ion velocity distribution function (VDF) moments, whereas Vlasiator provides the full VDFs. Appendix~\ref{app:training} contains more details on the training setup and computational complexity.

\begin{figure}[h]
    \centering
    \includegraphics[width=\textwidth]{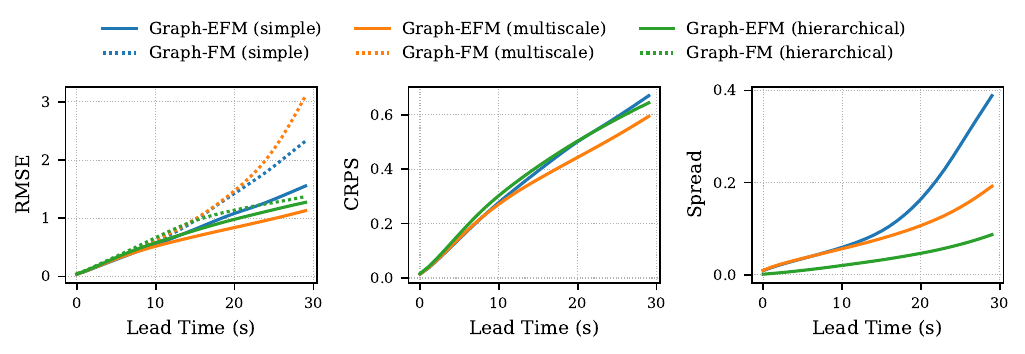}
    \caption{The mean of the normalized forecast RMSE, CRPS, and Spread over all variables.}
    \label{fig:mean_metrics}
\end{figure}

\paragraph{Results}
Figure~\ref{fig:example_forecast} displays a sample forecast from the hierarchical Graph-EFM, showing the ensemble mean and the ensemble standard deviation as a measure of forecast uncertainty. The performance of all models is summarized in Figure~\ref{fig:mean_metrics}, which plots the normalized RMSE, CRPS, and Spread averaged over all variables. To normalize the metrics as unitless scores, we divide each by the standard deviation of its corresponding variable. The RMSE measures the accuracy of the deterministic forecast (or the ensemble mean), while the CRPS provides a comprehensive assessment of the probabilistic forecast's accuracy and calibration. In the deterministic setting the hierarchical graph architecture accumulates less error than the simple and multiscale variations. We further find that probabilistic models achieve lower RMSE than their deterministic counterparts, particularly at longer lead times. We attribute this to the averaging over multiple ensemble members mitigating the impact of trajectory drift, resulting in a more stable estimate and less pronounced errors. The probabilistic models are overall quite comparable in performance. However, they tend to be underdispersed. In a well-calibrated forecasting system the ensemble spread should match the error~\cite{fortin2014should}, here the spread is smaller than the error. The limited sample size in our training data is most likely the main culprit here, but this is also a known characteristic of models trained primarily with a variational objective and in limited-area weather modeling~\cite{oskarsson2024probabilistic}, analogous to our magnetospheric setup in space. We also performed fairly conservative CRPS fine-tuning here, which is designed to alleviate underdispersion. Additional details on the metrics and per-variable results can be found in Appendices~\ref{app:metrics} and~\ref{app:results}.

\section{Conclusion and outlook}

In this work, we demonstrate that graph-based neural emulators can learn the complex dynamics of near-Earth space from a hybrid-Vlasov simulation, and provide exciting new uncertainty quantification capability for space weather prediction. Future development should focus on extending the model to three spatial dimensions, and depending on the simulation, a refined grid, for which GNNs are well suited. Training on a dataset with orders of magnitude more samples, spanning diverse real-world conditions across multiple solar cycles, would be ideal. This can also enable the use of larger timesteps, circumventing much of the cumulative error inherent to autoregressive emulators. Finally, looking into the physical realism of the forecasts and incorporating constraints, such as the divergence freeness of the magnetic field, is another promising future direction to follow.

\section*{Acknowledgements}

This work was funded by the Research Council of Finland under the FAISER project (grant nos. 361901, 361902). D.H. acknowledges support from the Fulbright-KAUTE Foundation Award for conducting research at UCSB. Computing resources were provided by the LUMI supercomputer, owned by the EuroHPC Joint Undertaking and hosted by CSC–IT Center for Science.

\section*{Code and data availability}

The source code for training and evaluating the machine learning models is available at \url{https://github.com/fmihpc/spacecast}. The dataset is produced using Vlasiator \url{https://github.com/fmihpc/vlasiator} and stored at \url{https://doi.org/10.5281/zenodo.16930055}.

\bibliographystyle{unsrt}  
\bibliography{references}


\clearpage

\appendix

\section{Future work}

\paragraph{Emulating kinetic-scale physics}
This work emulates the spatiotemporal evolution of velocity space moments of the ion VDFs, which Vlasiator solves fully. A natural progression would be to forecast the entire VDFs over time, rather than just their moments. Here one could take inspiration from previous work on emulating gyrokinetic simulations~\citeapp{galletti20255d}, where a hierarchical vision transformer was used for next step prediction of the 5D distribution function for adiabatic electrons.

\paragraph{Addressing error accumulation} Neural emulators struggle with error accumulation on long rollouts, causing simulated trajectories to diverge. This happens because the model increasingly operates ``out of distribution'' with each autoregressive step. To combat this, techniques like thermalization can significantly extend stable rollout times by controlling error growth \citeapp{pedersen2025thermalizer}. Another promising method to tackle cumulative error involves training models to forecast a future state in a single step while ensuring the temporal consistency of the forecast for each ensemble member~\citeapp{andrae2024continuous}.

\paragraph{Mesh refinement}
Simulating with hybrid-Vlasov models in three dimensions is computationally so expensive that mesh refinement becomes essential. This allows for higher spatial resolution in critical regions, like the bow shock and the magnetotail reconnection site near Earth, while using lower resolution in less important areas such as inflow and outflow boundaries \citeapp{ganse2023enabling}. This results in irregular data structures, making GNNs a future-proof model for modeling such data. However, recently physics-motivated adaptive refinement has been tested~\citeapp{kotipalo2024physics}, where the refinement itself is driven by physical conditions. This presents an additional interesting challenge for neural emulators.

\paragraph{Foundation models}
Recent community initiatives are gathering diverse physics simulations to create large-scale datasets for machine learning~\citeapp{angeloudi2024multimodal, ohana2024well}. Such datasets can be used to train foundation models, which can learn across scales or multiple physical systems~\citeapp{mccabe2024multiple, majid2024solaris, bodnar2025foundation, roy2025surya}. Space weather data would be a valuable addition to these growing datasets. Conversely, space weather prediction models could potentially be improved by training on diverse simulation data, allowing them to leverage insights from various physical domains and potentially generalize better.

\section{Dataset details}
\label{app:data}

The data used for this study is simulated using Vlasiator and contains the variables listed in Table~\ref{tab:dataset}.

\begin{table}[ht]
\centering
\caption{Summary of all variables and static fields in the Vlasiator dataset.}
\label{tab:dataset}
\begin{tabular}{lcc}
\toprule
 & \textbf{Abbreviation} & \textbf{Unit} \\
\midrule
\textbf{Variables} & & \\
\midrule
 Magnetic field $x$-component & $B_x$ & nT \\
Magnetic field $y$-component & $B_y$ & nT \\
Magnetic field $z$-component & $B_z$ & nT \\
Electric field $x$-component & $E_x$ & mV/m \\
Electric field $y$-component & $E_y$ & mV/m \\
Electric field $z$-component & $E_z$ & mV/m \\
Velocity field $x$-component & $v_x$ & km/s \\
Velocity field $y$-component & $v_y$ & km/s \\
Velocity field $z$-component & $v_z$ & km/s \\
Particle number density & $\rho$ & 1/cm$^{3}$ \\
Plasma temperature & $T$ & MK \\
Plasma pressure & $P$ & nPa \\
\midrule
\textbf{Static fields} & & \\
\midrule
Coordinate in $x$ & $x$ & $R_E$ \\
Coordinate in $z$ & $z$ & $R_E$ \\
Radial distance from Earth & $r$ & $R_E$ \\
\bottomrule
\end{tabular}
\end{table}

\section{Model details}
\label{app:model}

\begin{figure}[htbp]
    \centering
    \begin{subfigure}[b]{0.48\textwidth}
        \centering
        \includegraphics[width=\textwidth]{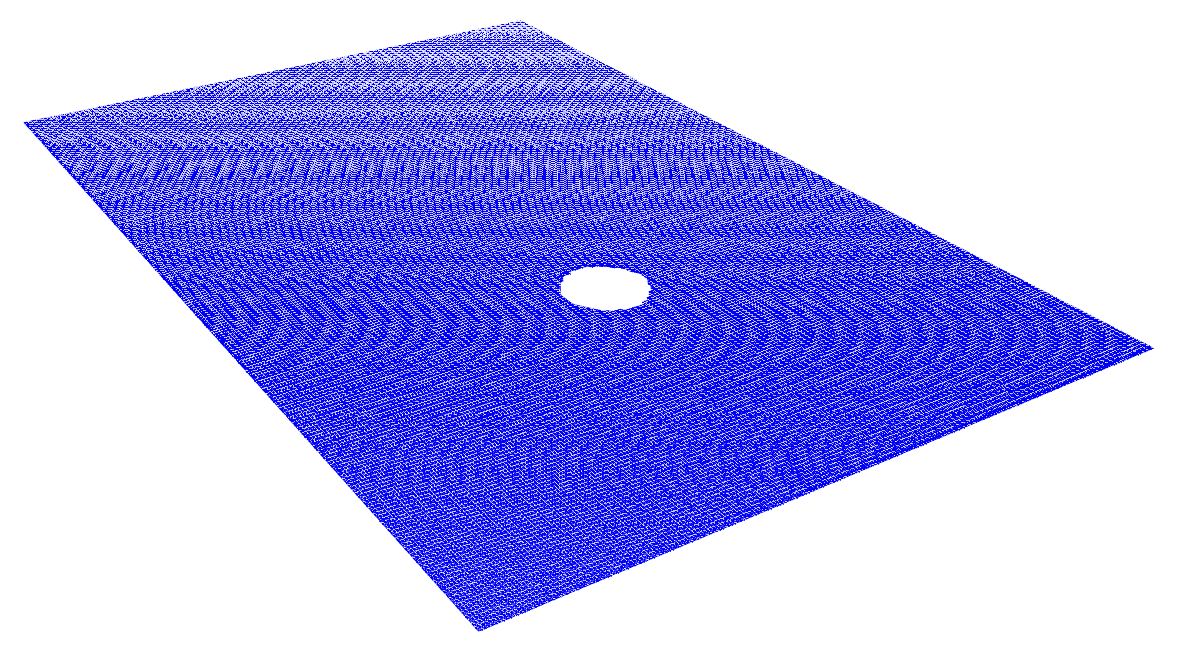}
        \caption{Simple Graph}
        \label{fig:simple}
    \end{subfigure}
    \hfill
    \begin{subfigure}[b]{0.48\textwidth}
        \centering
        \includegraphics[width=\textwidth]{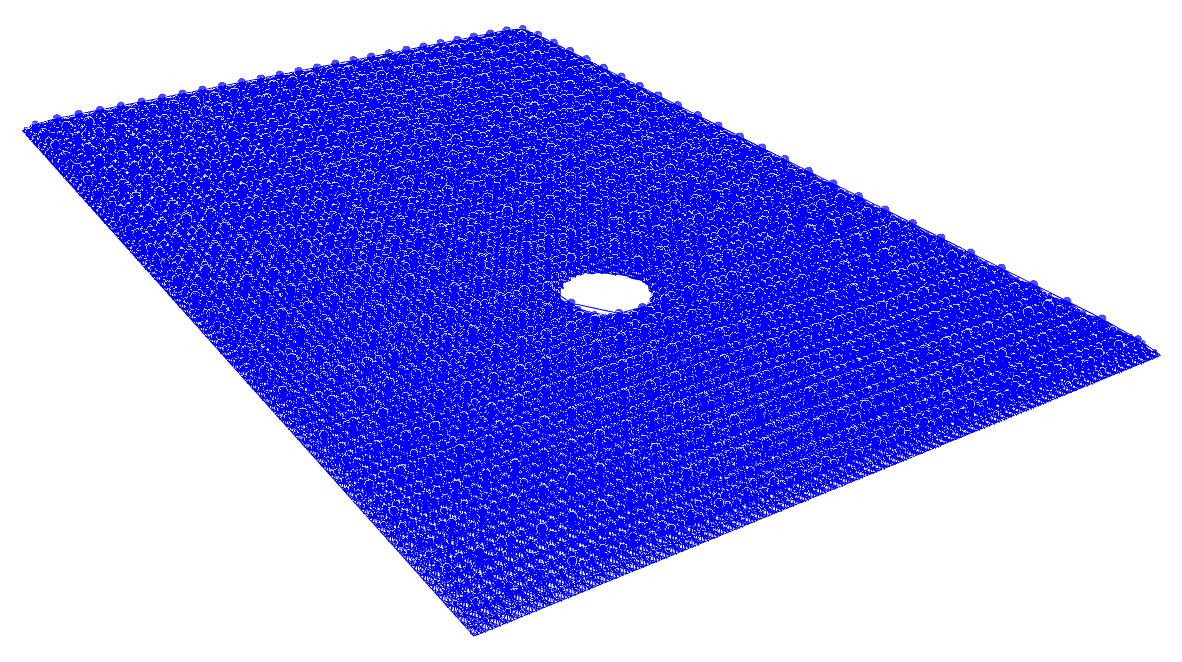}
        \caption{Multiscale Graph}
        \label{fig:multiscale}
    \end{subfigure}
    \vspace{1em}
    \begin{subfigure}[b]{0.48\textwidth}
        \centering
        \includegraphics[width=\textwidth]{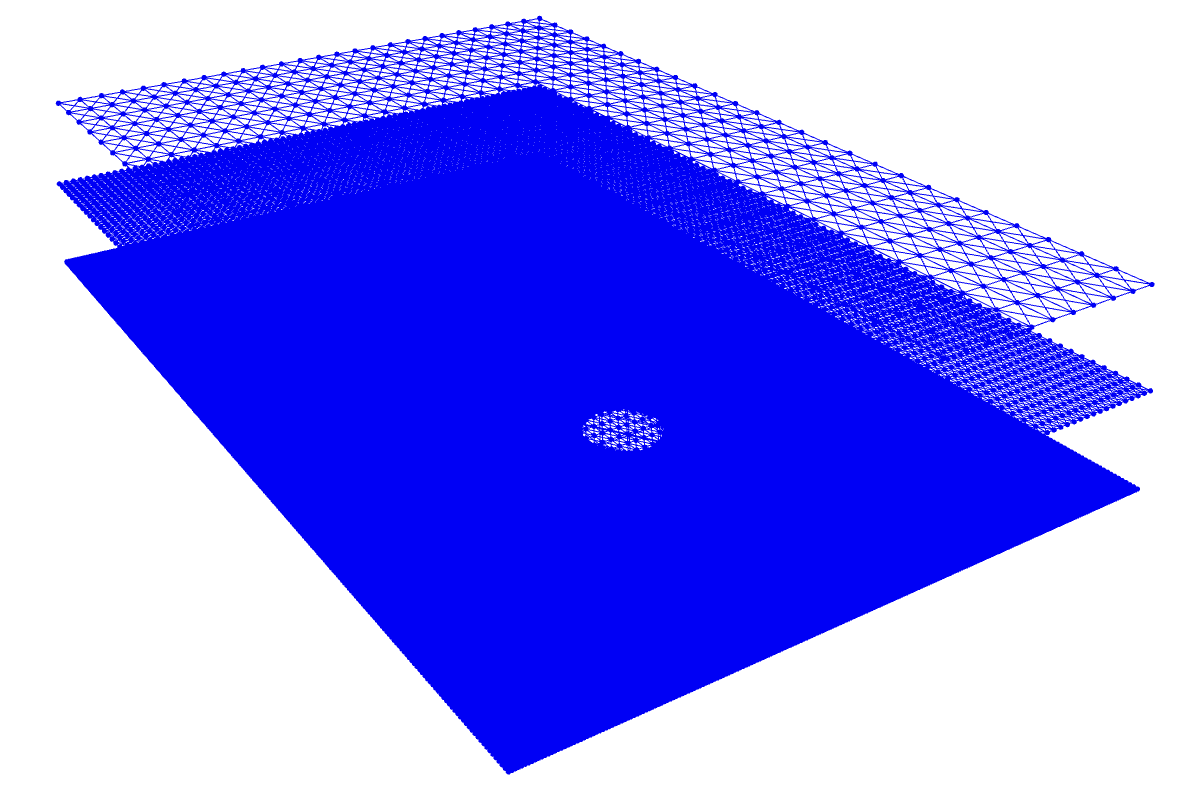}
        \caption{Graph Layers}
        \label{fig:layers}
    \end{subfigure}
    \hfill
    \begin{subfigure}[b]{0.48\textwidth}
        \centering
        \includegraphics[width=\textwidth]{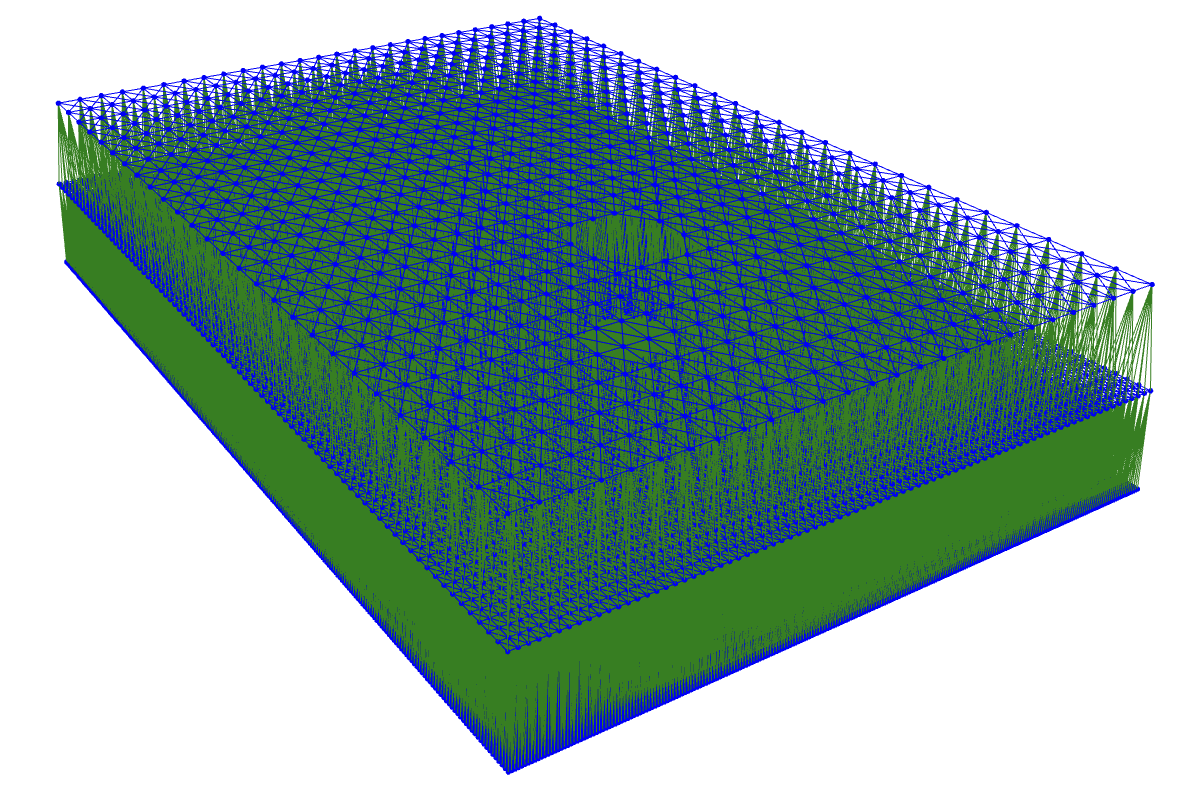}
        \caption{Hierarchical Graph}
        \label{fig:hierarchical}
    \end{subfigure}

    \caption{Quadrilateral mesh variations used by the GNNs. Nodes and same-level edges in blue, and inter-level edges in green. Node size corresponds to degree in the multiscale graph.}
    \label{fig:graphs}
\end{figure}

The $671\times1006\ (z, x)$ data grid excludes $5124$ inner boundary nodes. We construct graphs by recursively downsampling the grid, placing each coarser-level node at the center of a $3\times3$ finer-level node square as seen in Fig.~\ref{fig:graphs}. We compare three mesh architectures: a simple single-level graph, a three-level multiscale graph, and a three-level hierarchical graph, with full statistics in Table~\ref{tab:graphs}. The multiscale version connects all levels into a single heterogeneous mesh to capture both short- and long-range interactions, while the hierarchical version uses distinct, uniform levels to process information at different spatial scales separately. For each node, an MLP encodes static features in Table~\ref{tab:dataset}, concatenated with previous states. For a complete explanation on update rules in the GNNs see~\cite{oskarsson2024probabilistic}. All MLPs use one hidden layer with Swish activation~\citeapp{ramachandran2017swish} and layer normalization~\citeapp{ba2016layernorm}.

\begin{table}[h]
\centering
\caption{Number of nodes and edges in the different graphs.}
\begin{tabular}{llrr}
\toprule
\textbf{Graph} & & \textbf{Nodes} & \textbf{Edges} \\
\midrule
Simple & $\mathcal{G}_\mathrm{S}$ & 58592 & 465584 \\
\midrule
Multiscale & $\mathcal{G}_\mathrm{MS}$ & 58592 & 522054 \\
\midrule
\multirow{6}{*}{Hierarchical} & $\mathcal{G}_0$         & 58592     & 465584 \\
& $\mathcal{G}_{0, 1} / \mathcal{G}_{1,0}$ & -        & 58592  \\
& $\mathcal{G}_1$         & 6510      & 51032    \\
& $\mathcal{G}_{1, 2} / \mathcal{G}_{2, 1}$ & -        & 6510   \\
& $\mathcal{G}_2$         & 723       & 5438 \\
\cmidrule{2-4}
& Total & 65825    & 587156  \\
\midrule
$\mathcal{G}_\text{G2M}$ &     & -       & 1411687 \\
$\mathcal{G}_\text{M2G}$ &     & -       & 2679608 \\
\midrule
Grid & & 669902   & - \\
\bottomrule
\end{tabular}
\label{tab:graphs}
\end{table}

\clearpage

\paragraph{Deterministic objective}

We train the deterministic models by minimizing a weighted MSE:
\begin{equation}
\mathcal{L} = \frac{1}{N} \sum_{n=1}^N \sum_{i=1}^{d_x} \omega_i \lambda_i \left( \hat{X}_{n, i} - X_{n, i} \right)^2.
\label{eq:mse}
\end{equation}
The loss is weighted by two terms: $\lambda_i$ is the inverse variance of the time differences for variable $j$, which normalizes the contribution of variables with different dynamic ranges. The second term, $\omega_i$, is a variable-specific weight. While often used in atmospheric forecasting to prioritize variables by altitude, we set it uniformly to $1/d_x$.

\paragraph{Probabilistic objective}

The probabilistic model's single-step prediction has a structure analogous to a conditional VAE, and it is trained by optimizing a variational objective derived from the ELBO:
\begin{align}
    \begin{split}
    \label{eq:elbo_loss}
    \varloss&\left(X^{t-2:t-1}, X^t \right)
    =
    \lambda_{\mathrm{KL}} D_{\mathrm{KL}}\left({\vardist }\middle|\middle|\,{\priordist}\right)
    \\
    -&\mathbb{E}_{\vardist}\left[{
        {\textstyle \sum_{n=1}^N
        \sum_{i = 1}^{d_x}}
        \log \mathcal{N}\left({
            X^{t}_{v, i}
        }\middle|{
            g\left(Z^t, \arcond\right)_{v, i}
        },{
            \sigma^2_{v, i}
        }\right)
    }\right].
    \end{split}
\end{align}
This objective consists of two terms. The first is the KL divergence, which acts as a regularizer, encouraging the approximate posterior distribution $q$ to remain close to the prior $p$. The second term is the expected negative log-likelihood, or reconstruction loss, which trains the predictor $g$ to accurately reconstruct the true state $X^t$ from a latent sample $Z^t$. The hyperparameter $\lambda_\mathrm{KL}$ balances these two terms to prevent the model from collapsing to a deterministic prediction. As both distributions are Gaussian, the KL divergence has a closed-form solution. The variance $\sigma^2_{n, i}$ is an output of the predictor network. The model is further fine-tuned using the CRPS loss, which is minimized only when the predicted distribution matches the empirical data distribution. We use an unbiased two-sample estimator for the CRPS loss, summed over all grid points and variables:
\begin{equation}
    \label{eq:crps_loss}
    \crpsloss = \sum_{n=1}^N \sum_{i = 1}^{d_x}
    \frac{1}{2} \left(
        \left|\hat{X}_{n , i} - X_{n, i}\right|
        + \left|\check{X}_{n, i} - X_{n, i}\right|
        - \left|\hat{X}_{n, i} - \check{X}_{n, i}\right|
    \right)
\end{equation}
where $\hat{X}^t$ and $\check{X}^t$ are two independent ensemble members (forecasts) generated by the model. The final training loss combines both objectives: $\mathcal{L} = \varloss + \lambda_\mathrm{CRPS} \crpsloss$.

\section{Training details}
\label{app:training}

All models are trained using the AdamW optimizer~\citeapp{loshchilov2019adamw} with parameters $\beta_1=0.9$, $\beta_2=0.95$, a weight decay of $0.01$, and an effective batch size of 8. The specific training schedules for the deterministic and probabilistic models are detailed in Table~\ref{tab:schedule}. The deterministic Graph-FM model follows a standard training schedule with a learning rate decay after 300 epochs. The probabilistic Graph-EFM model is trained in three distinct stages. First, we train the model as a simple autoencoder with both $\lambda_\text{KL}$ and $\lambda_\text{CRPS}$ set to $0$ to teach the encoder and decoder to reconstruct the fields from the latent space without probabilistic constraints. In the second stage, we introduce the KL divergence term to train the full variational model on the single-step distribution. Finally, we fine-tune the model with the CRPS loss to improve the calibration of the ensemble.

\begin{table}[htbp]
\centering
\caption{Training schedules for the deterministic and probabilistic models.}
\label{tab:schedule}
\begin{tabular}{lcccc}
\toprule
\textbf{Model} & \textbf{Epochs} & \textbf{Learning Rate} & \textbf{$\lambda_\text{KL}$} & \textbf{$\lambda_\text{CRPS}$} \\
\midrule
\multirow{2}{*}{Graph-FM} & 300 & $10^{-3}$ & - & - \\
 & 200 & $10^{-4}$ & - & - \\
\midrule
\multirow{3}{*}{Graph-EFM} & 200 & $10^{-3}$ & 0 & 0 \\
 & 300 & $10^{-3}$ & 1 & 0 \\
 & 50 & $10^{-4}$ & 1 & $10^{-3}$ \\
\bottomrule
\end{tabular}
\end{table}

\paragraph{Computational complexity}

Table~\ref{tab:model_times} shows the times it took to train the different models, and their inference speeds on the test set. Training was done on 8 AMD MI250X GPUs with 2 workers for the dataloader. The actual training time is therefore the GPU hours in the table divided by 8. Inference was performed on 1 AMD MI250X GPU and is measured as the time it takes for one next-step prediction. Graph-EFM produces an ensemble size of 5 here. For comparison, the Vlasiator simulation takes 4--5 minutes on 50 AMD EPYC 7H12 CPUs to simulate 1\,s of real time.
\begin{table}[h!]
\centering
\caption{Training and inference times for different models and graph variations.}
\label{tab:model_times}

\begin{tabular}{llcc}
\toprule
\textbf{Model} & \textbf{Graph} & \textbf{Training time (GPU h)} & \textbf{Inference time (s)} \\
\midrule
\multirow{3}{*}{Graph-FM} & Simple & 101 & 0.47 \\
 & Multiscale & 102 & 0.48 \\
 & Hierarchical & 108 & 0.52 \\
\midrule

\multirow{3}{*}{Graph-EFM} & Simple & 119 & 3.20 \\
 & Multiscale & 122 & 3.31 \\
 & Hierarchical & 131 & 3.45 \\
\bottomrule
\end{tabular}
\end{table}

\section{Metrics}
\label{app:metrics}

To evaluate the performance of our forecasts, we use a set of standard metrics. For an ensemble forecast with $M$ members (deterministic forecasts have $M=1$), we denote the prediction for variable $i$ at location $n$ for a given sample $s$ at time $t$ as $\hat{X}_{n,i}^{s,t,m}$, with the corresponding ground truth being $X_{n,i}^{s,t}$. RMSE is calculated by averaging the squared error over all $S$ forecasts in the test set and all $N$ grid locations:
\begin{align}
\mathrm{RMSE}_{t,i} &= \sqrt{\frac{1}{SN}\sum_{s=1}^{S}\sum_{n=1}^N \left(\overline{X}_{n,i}^{s,t} - X_{n,i}^{s,t}\right)^{2}} \\
\text{where} \quad \overline{X}_{n,i}^{s,t} &= \frac{1}{M}\sum_{m=1}^{M}\hat{X}_{n,i}^{s,t,m}.
\end{align}

The CRPS is calculated for ensemble forecasts to assess the overall quality of a probabilistic forecast by comparing the entire predictive distribution to the single ground truth observation. Lower values indicate better performance. We use a finite-sample estimate~\citeapp{zamo2018estimation}, computed as:
\begin{align}
\begin{split}
\mathrm{CRPS}_{t,i} = \frac{1}{SN}\sum_{s=1}^{S}\sum_{n=1}^N \Biggl( &\frac{1}{M}\sum_{m=1}^{M}\left|\hat{X}_{n,i}^{s,t,m} - X_{n,i}^{s,t}\right| \\
&-\frac{1}{2M(M-1)}\sum_{m=1}^{M}\sum_{m'=1}^{M}\left|\hat{X}_{n,i}^{s,t,m} - \hat{X}_{n,i}^{s,t,m'}\right| \Biggr).
\end{split}
\end{align}

The spread quantifies the uncertainty expressed by the ensemble. It is defined as the root mean square deviation of the ensemble members from the ensemble mean.
\begin{equation}
\mathrm{Spread}_{t,i} = \sqrt{\frac{1}{SMN}\sum_{s=1}^{S}\sum_{m=1}^{M}\sum_{n=1}^N \left(\overline{X}_{n,i}^{s,t} - \hat{X}_{n,i}^{s,t,m}\right)^{2}}
\end{equation}

\section{Additional results}
\label{app:results}

\begin{figure}[h]
    \centering
    \includegraphics[width=\textwidth]{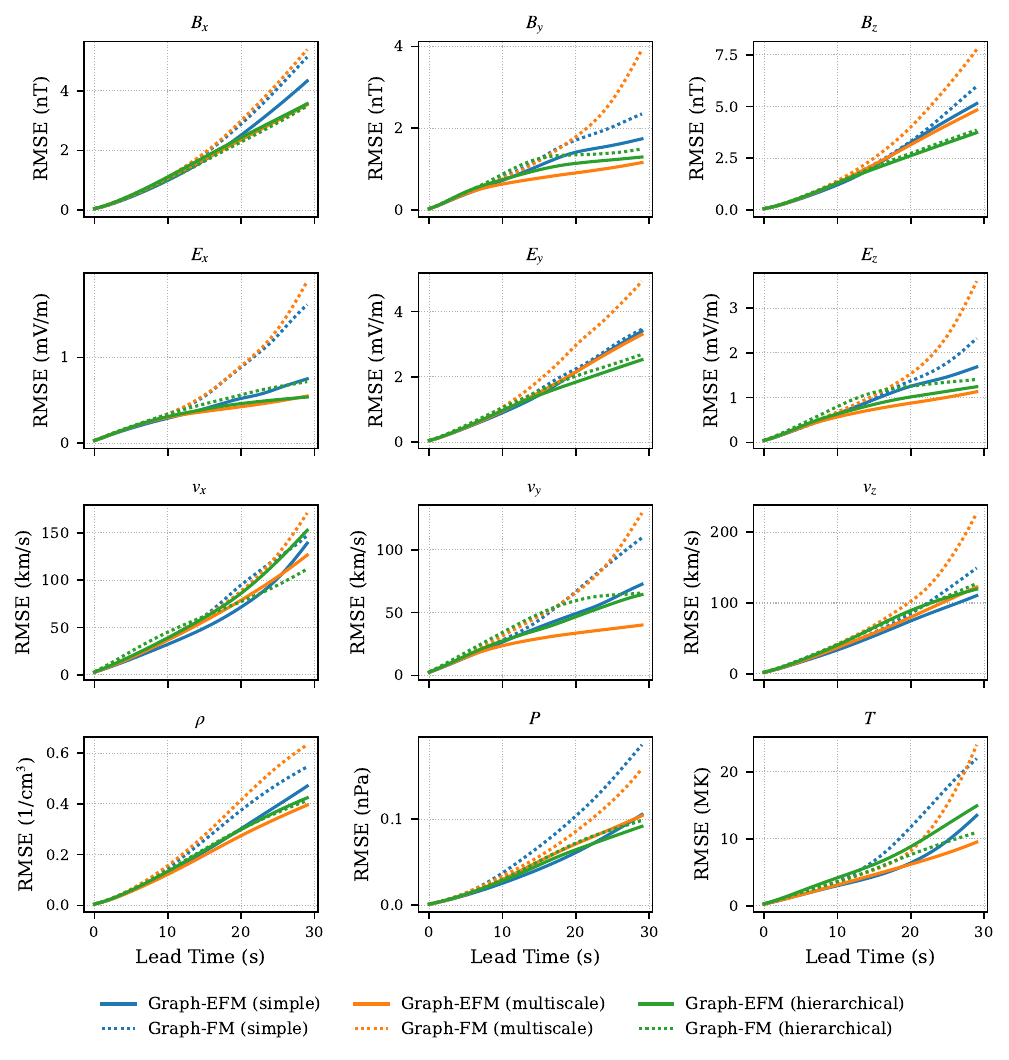}
    \caption{RMSE for all predicted variables as a function of forecast lead time.}
    \label{fig:rmse}
\end{figure}

\begin{figure}
    \centering
    \includegraphics[width=\textwidth]{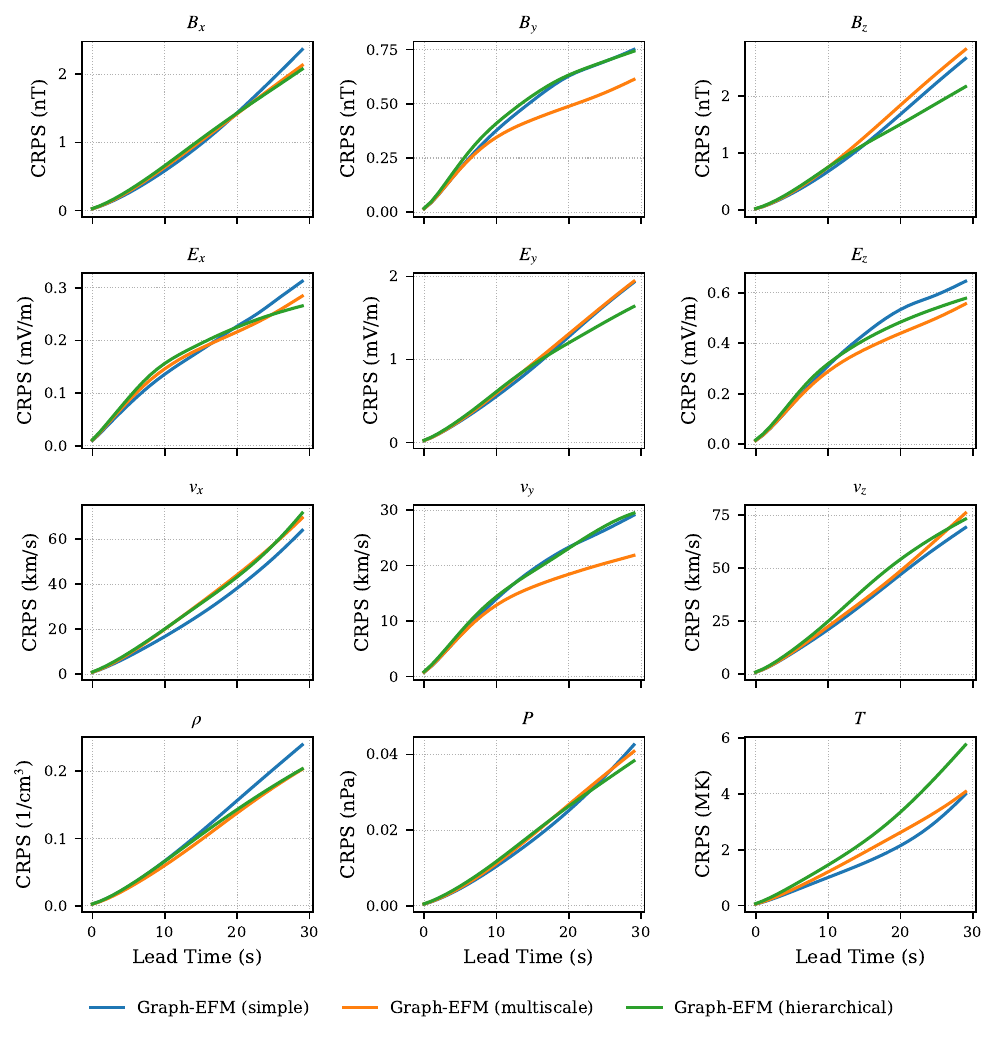}
    \caption{CRPS for all predicted variables as a function of forecast lead time.}
    \label{fig:crps}
\end{figure}

\begin{figure}
    \centering
    \includegraphics[width=\textwidth]{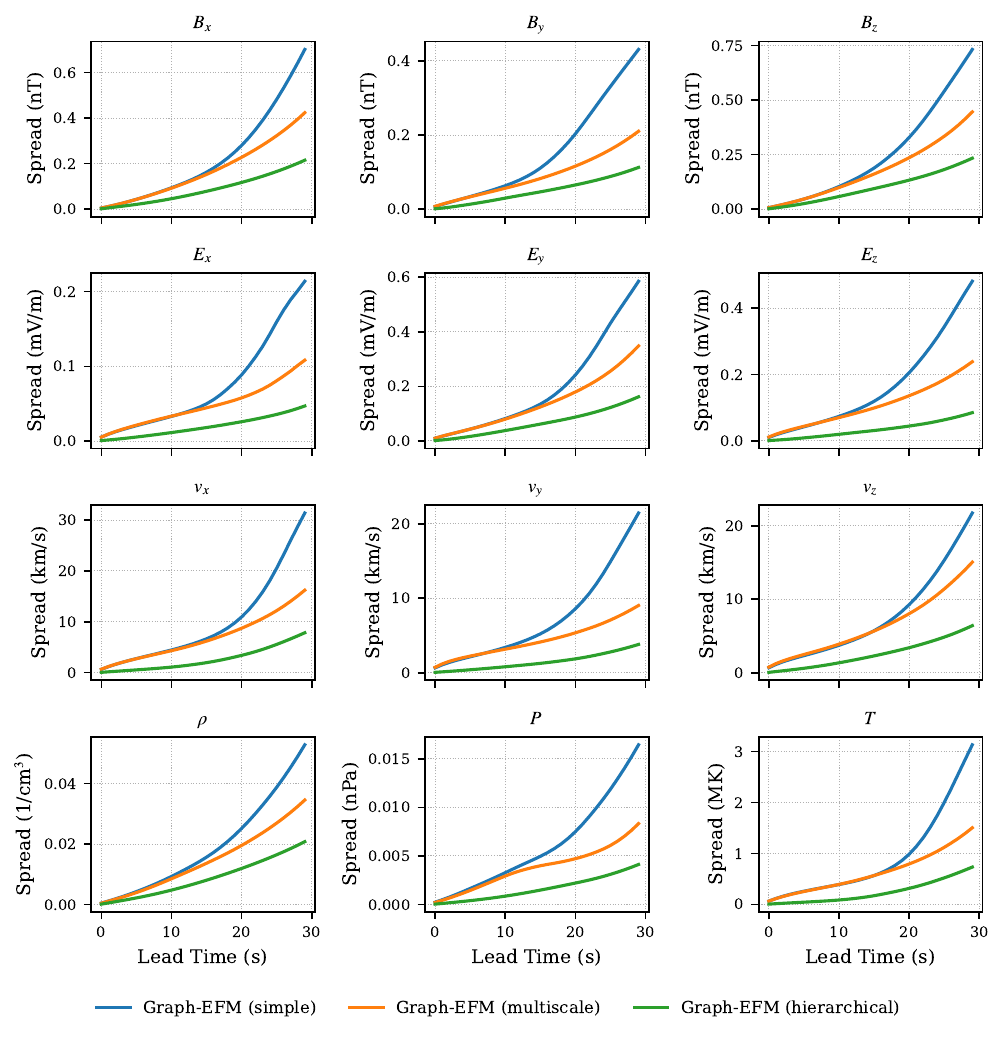}
    \caption{Ensemble spread for all predicted variables as a function of forecast lead time.}
    \label{fig:spread}
\end{figure}

Figure~\ref{fig:rmse} shows that for deterministic models, the error in the simple and multiscale architectures grows more rapidly than in the hierarchical version. The ensemble models perform similarly to each other in terms of their mean error. The CRPS scores in Fig.~\ref{fig:crps} are also quite comparable. The ensemble spread shown in Fig.~\ref{fig:spread} is consistently lower than the error for all models, indicating that the forecasts are underdispersed. When these two values are equal, the forecasting system is considered well-calibrated.

\begin{figure}[h]
    \centering
    \includegraphics[width=\textwidth]{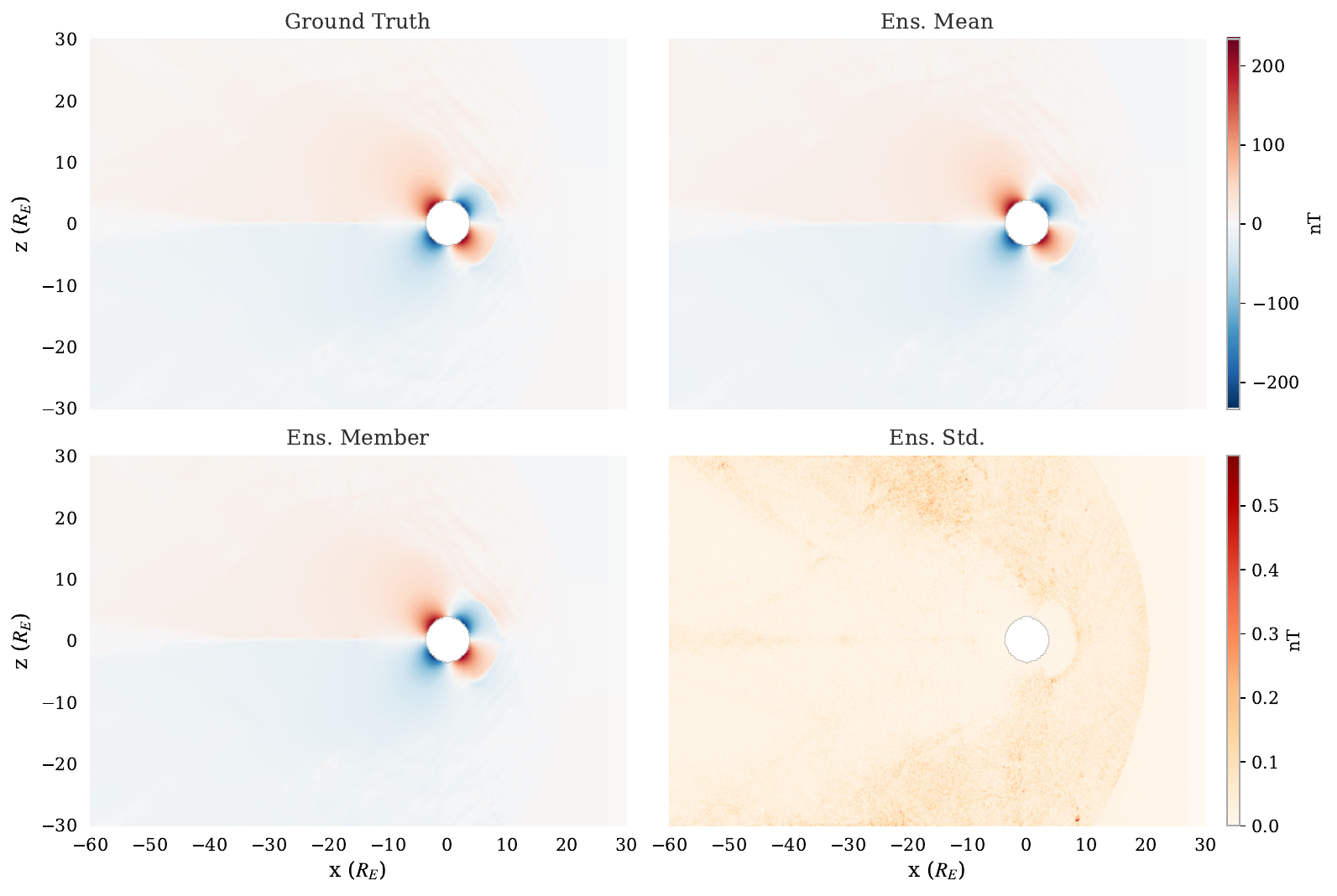}
    \caption{Magnetic field component $B_x$ at timestep 10 from Graph-EFM (hierarchical).}
    \label{fig:bx}
\end{figure}

\begin{figure}[h]
    \centering
    \includegraphics[width=\textwidth]{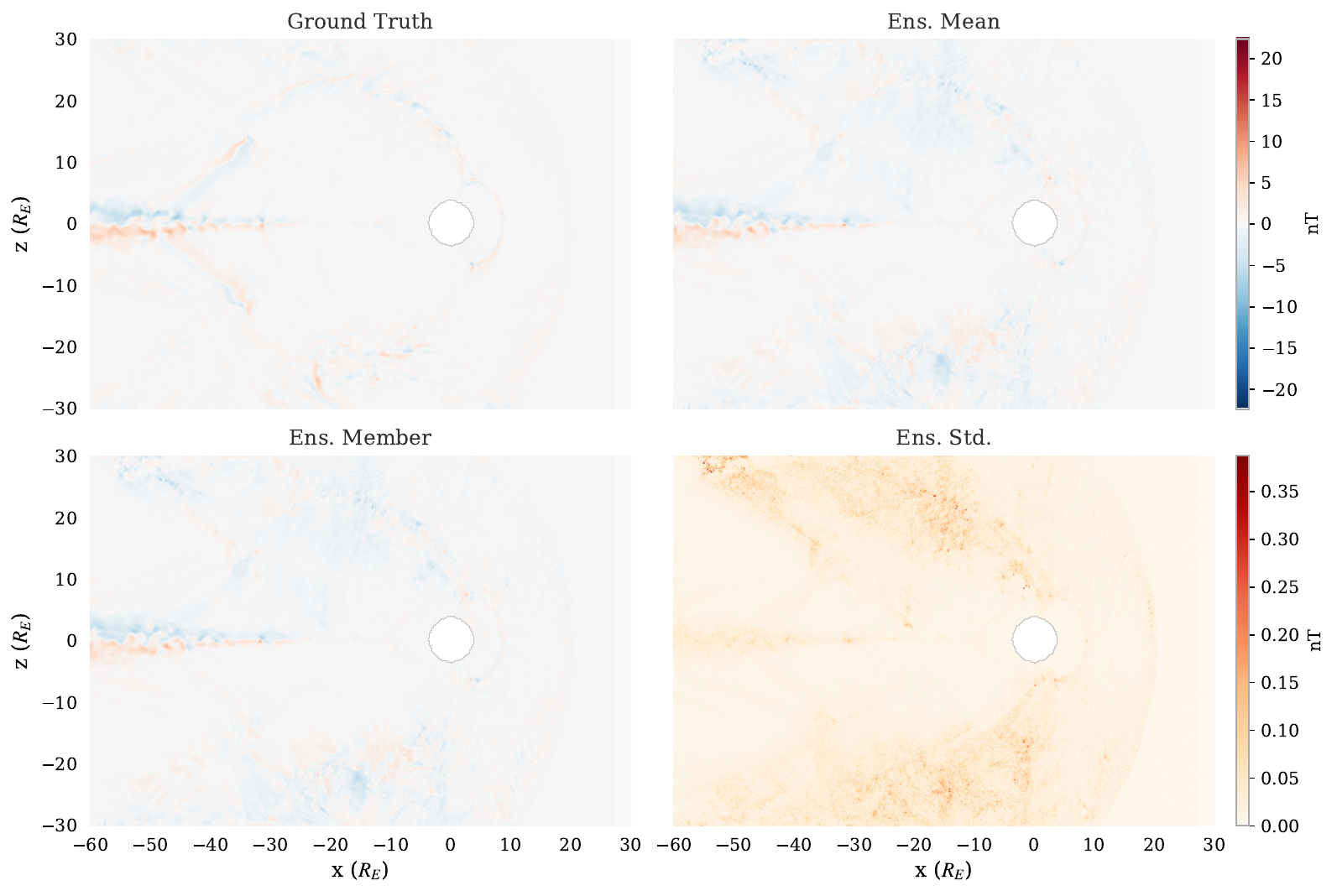}
    \caption{Magnetic field component $B_y$ at timestep 10 from Graph-EFM (hierarchical).}
    \label{fig:by}
\end{figure}

\begin{figure}[h]
    \centering
    \includegraphics[width=\textwidth]{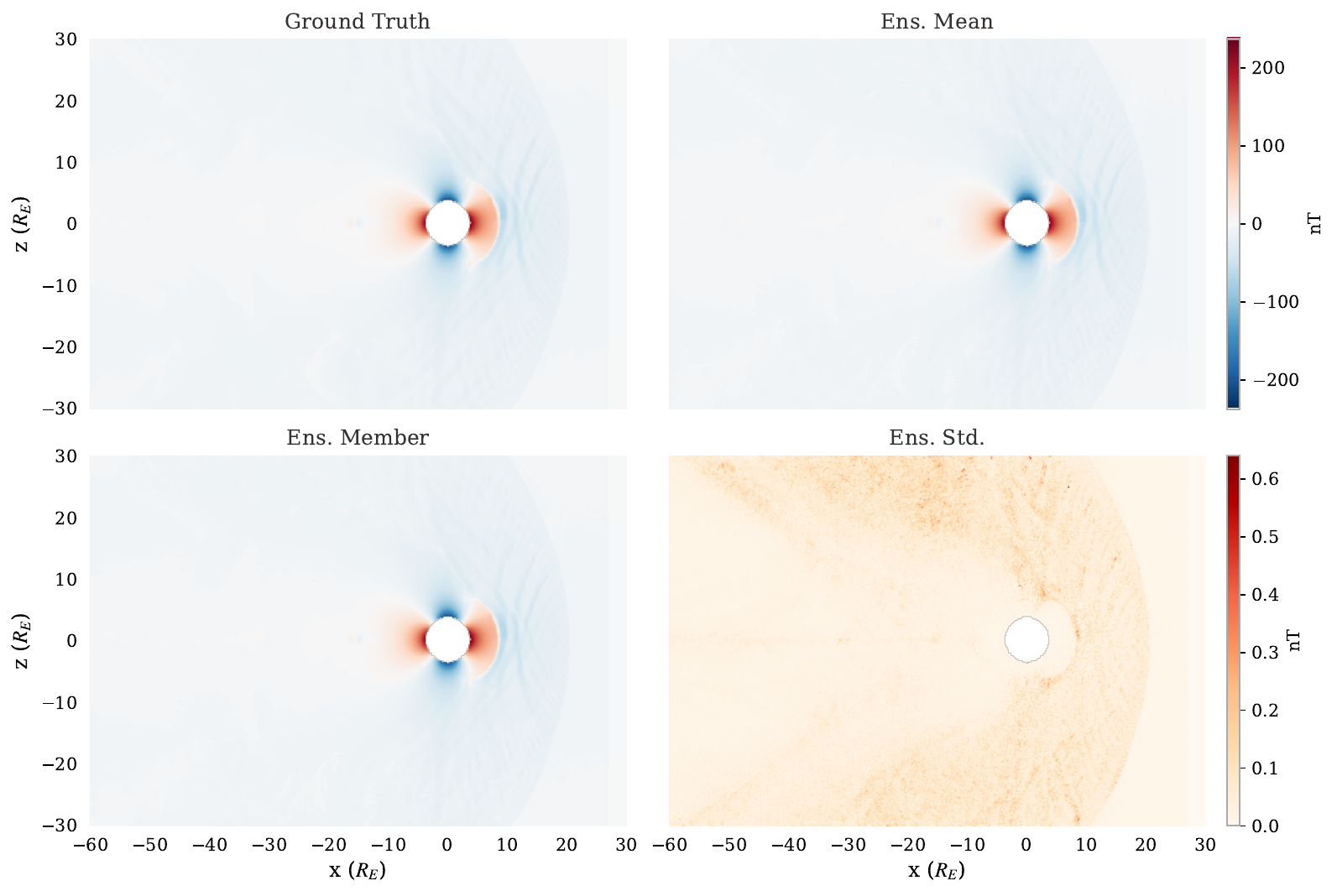}
    \caption{Magnetic field component $B_z$ at timestep 10 from Graph-EFM (hierarchical).}
    \label{fig:bz}
\end{figure}

\begin{figure}[h]
    \centering
    \includegraphics[width=\textwidth]{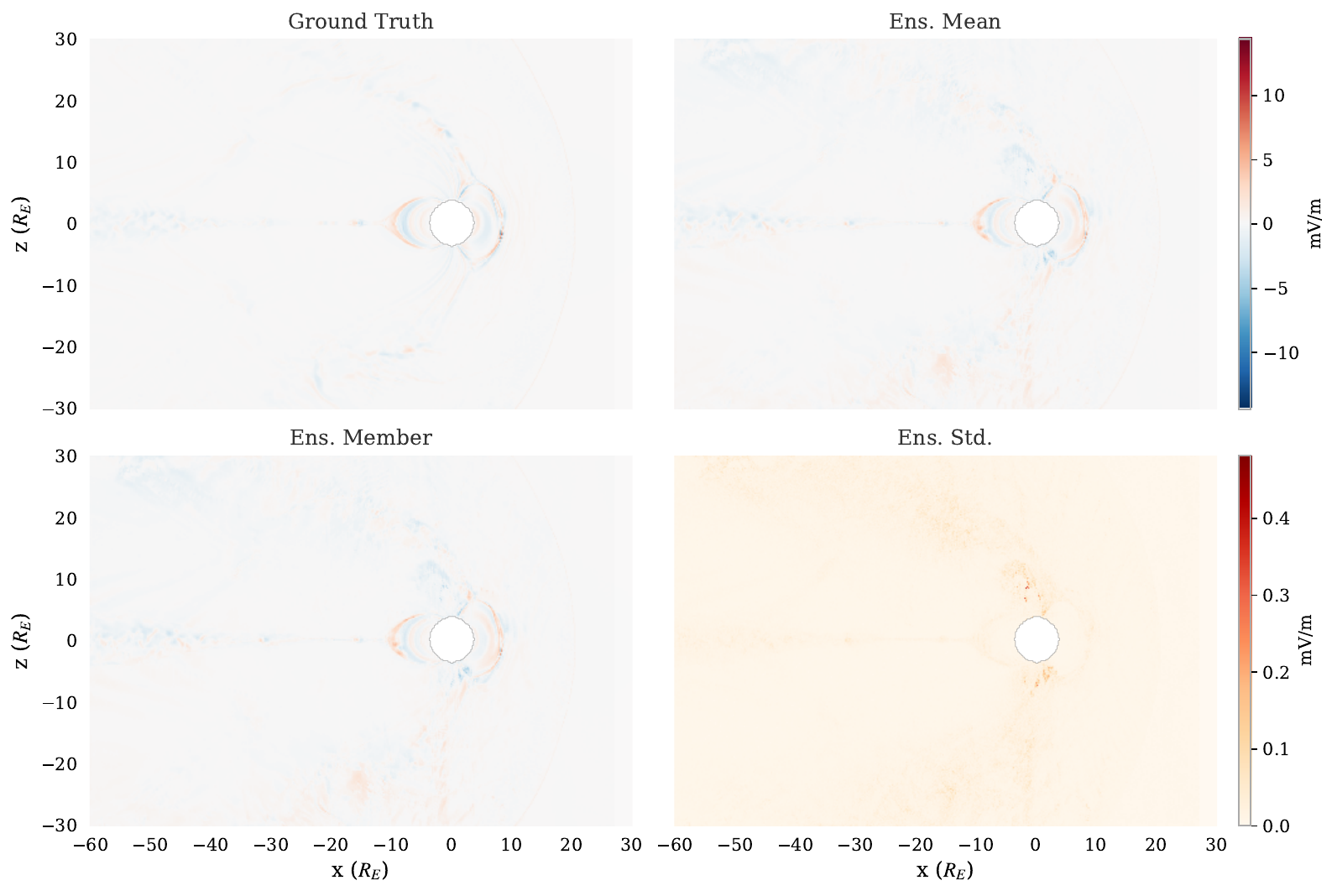}
    \caption{Electric field component $E_x$ at timestep 10 from Graph-EFM (hierarchical).}
    \label{fig:ex}
\end{figure}

\begin{figure}[h]
    \centering
    \includegraphics[width=\textwidth]{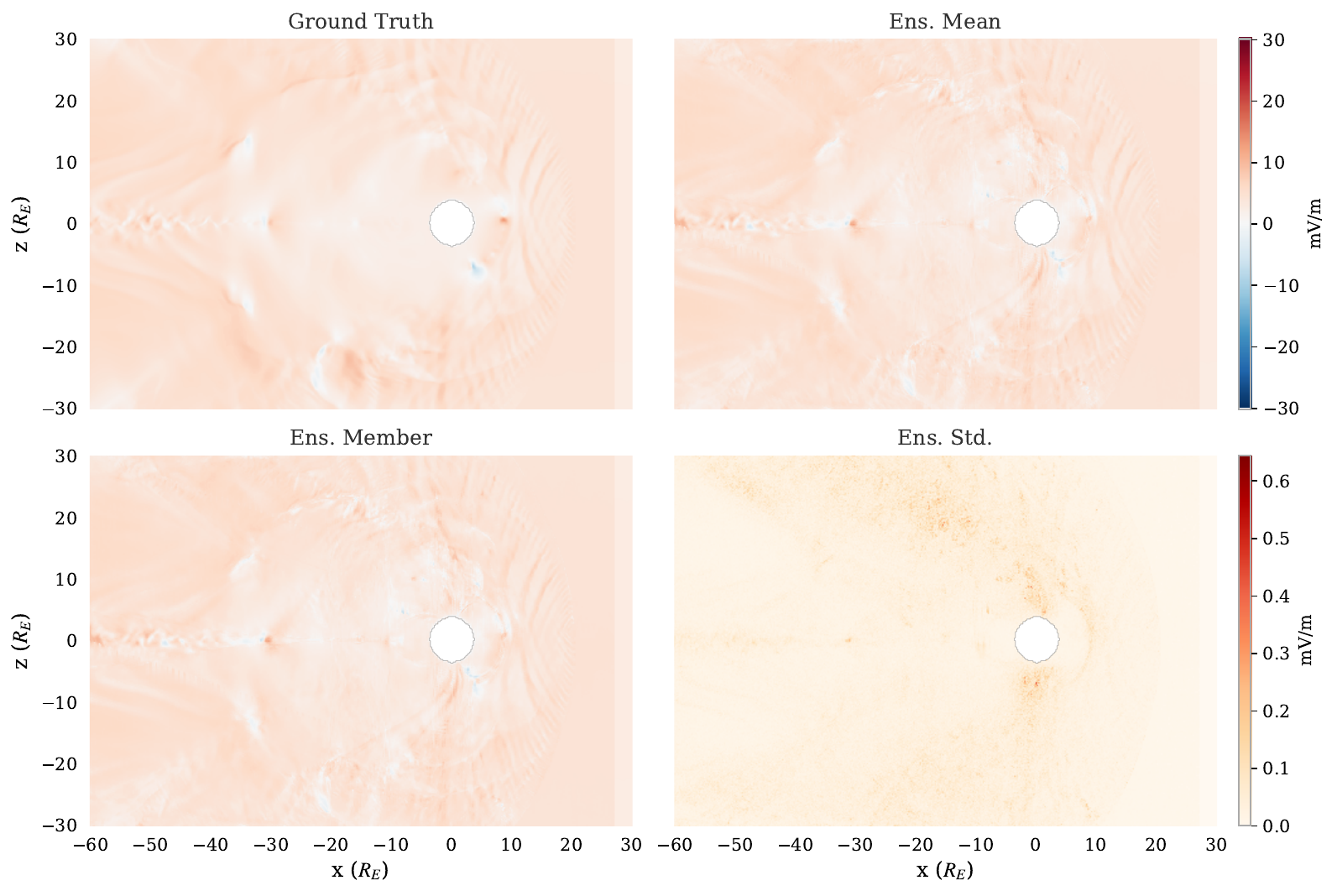}
    \caption{Electric field component $E_y$ at timestep 10 from Graph-EFM (hierarchical).}
    \label{fig:ey}
\end{figure}

\begin{figure}[h]
    \centering
    \includegraphics[width=\textwidth]{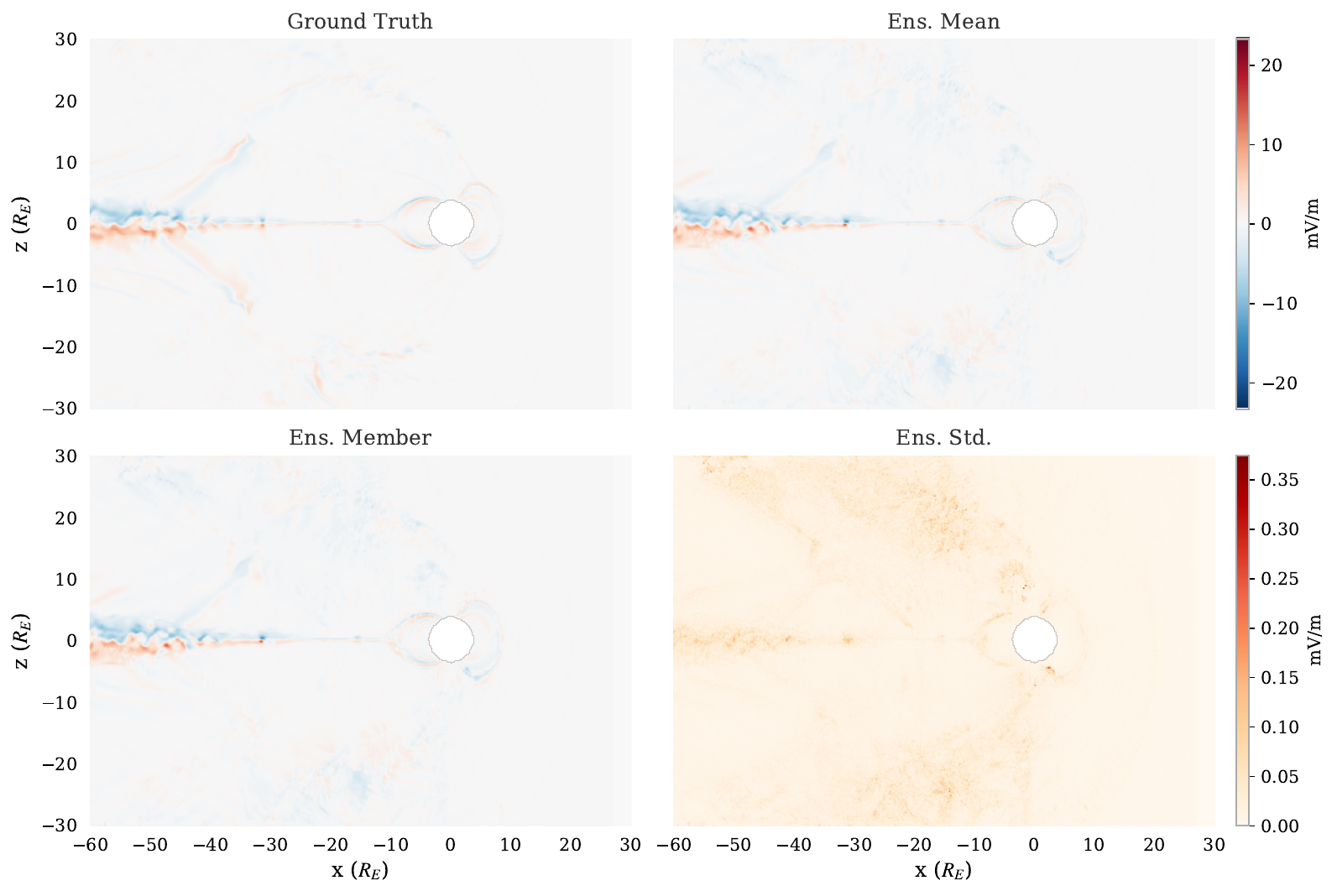}
    \caption{Electric field component $E_z$ at timestep 10 from Graph-EFM (hierarchical).}
    \label{fig:ez}
\end{figure}

\begin{figure}[h]
    \centering
    \includegraphics[width=\textwidth]{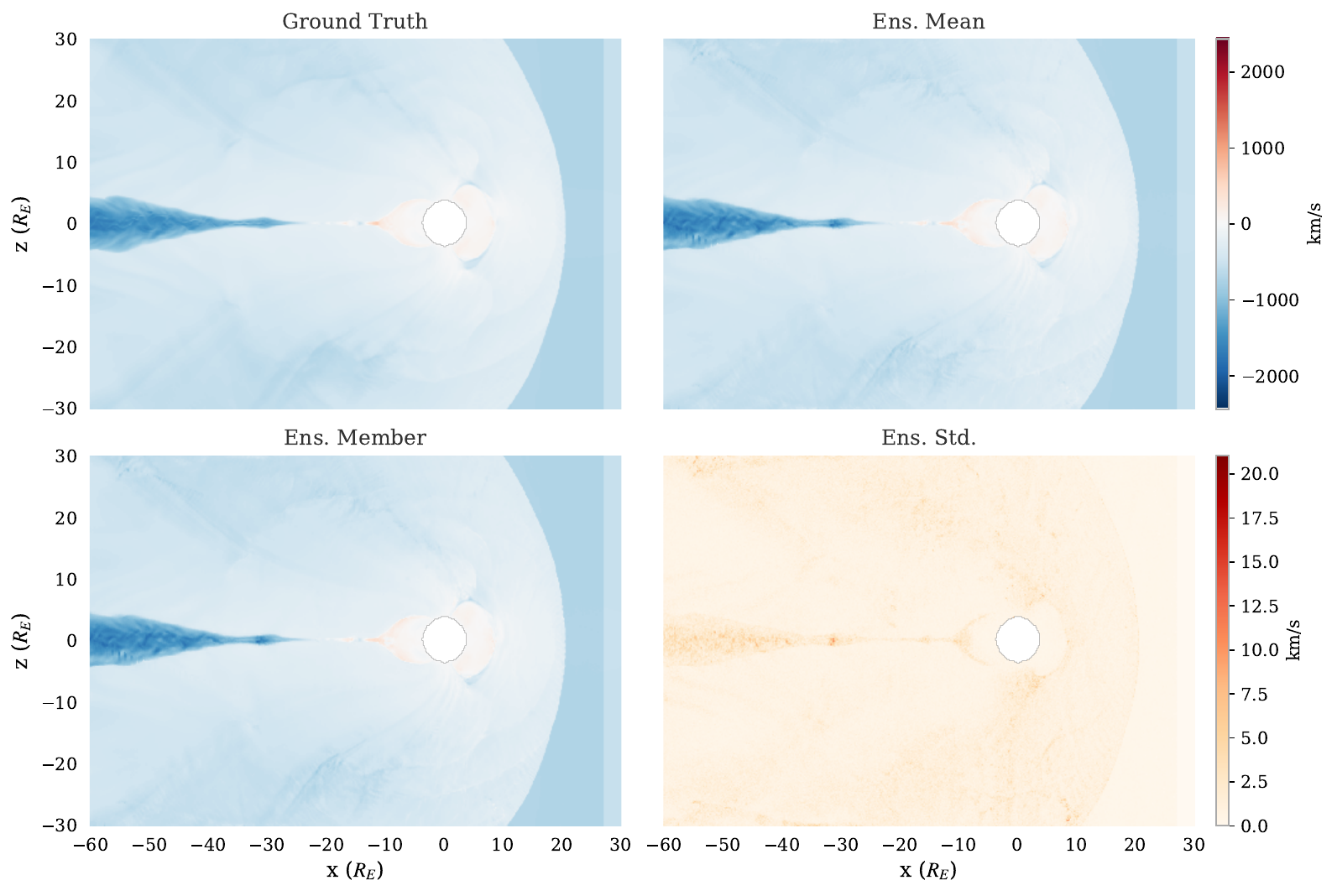}
    \caption{Velocity component $v_x$ at timestep 10 from Graph-EFM (hierarchical).}
    \label{fig:vx}
\end{figure}

\begin{figure}[h]
    \centering
    \includegraphics[width=\textwidth]{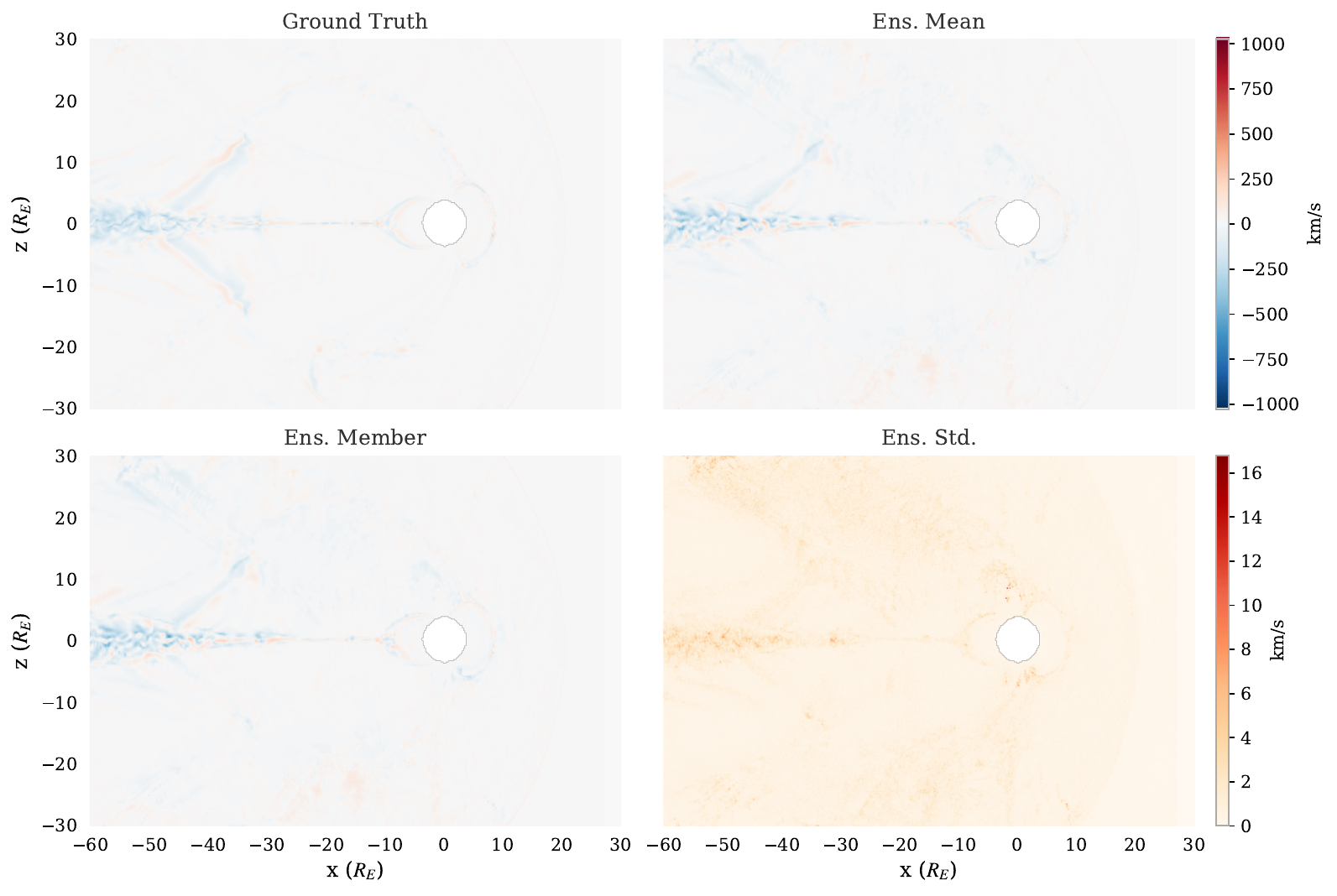}
    \caption{Velocity component $v_y$ at timestep 10 from Graph-EFM (hierarchical).}
    \label{fig:vy}
\end{figure}

\begin{figure}[h]
    \centering
    \includegraphics[width=\textwidth]{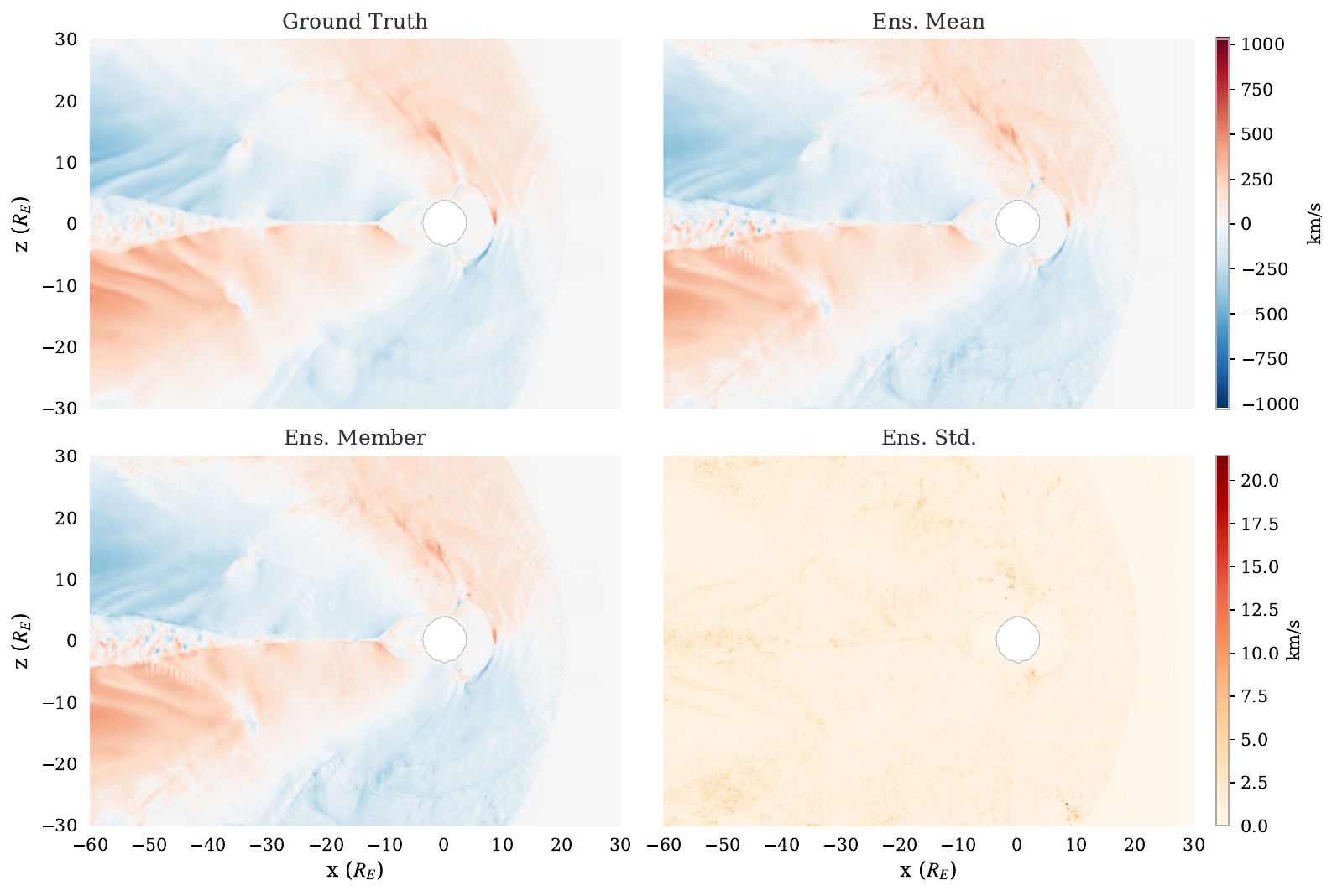}
    \caption{Velocity component $v_z$ at timestep 10 from Graph-EFM (hierarchical).}
    \label{fig:vz}
\end{figure}

\begin{figure}[h]
    \centering
    \includegraphics[width=\textwidth]{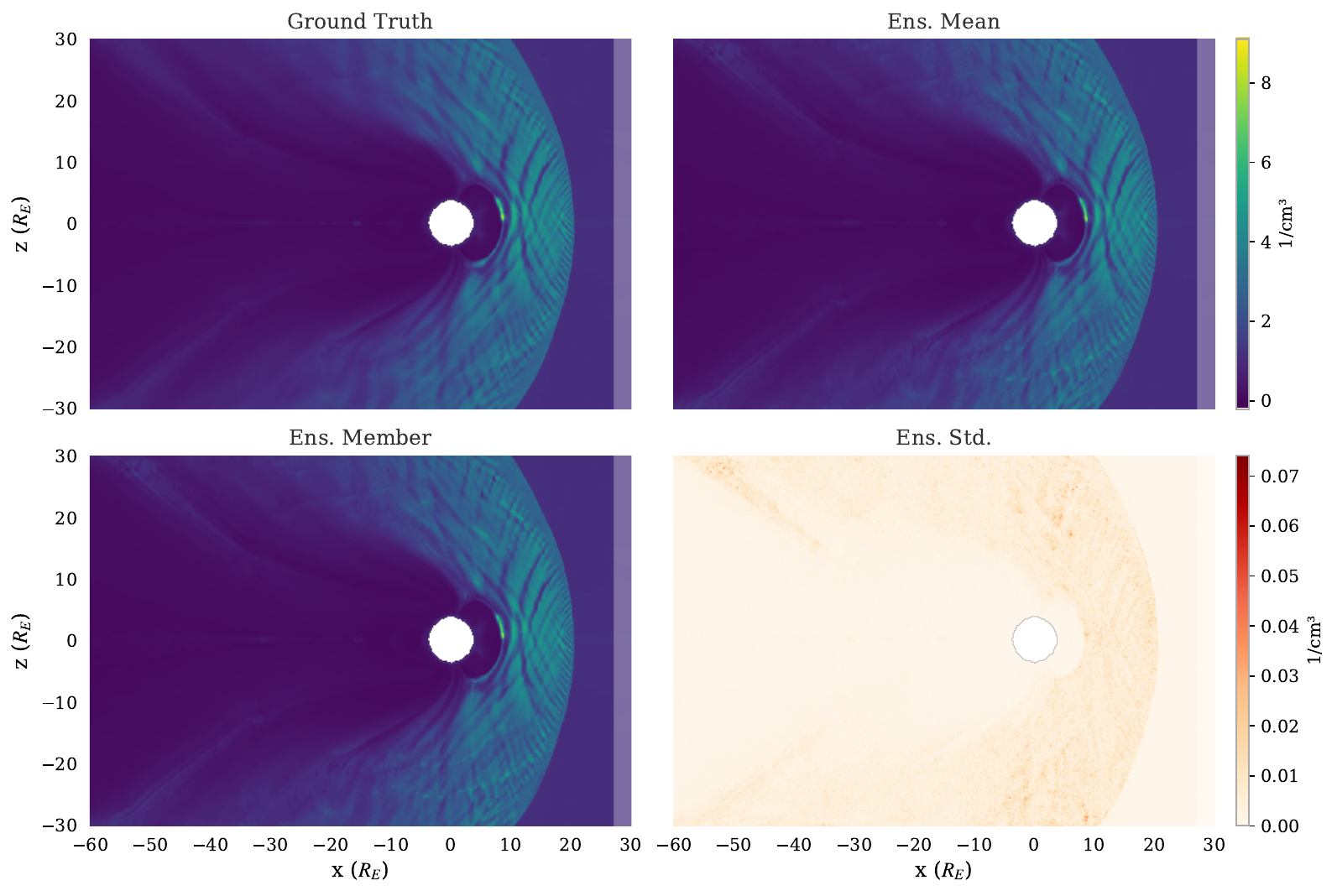}
    \caption{Particle number density $\rho$ at timestep 10 from Graph-EFM (hierarchical).}
    \label{fig:rho}
\end{figure}

\begin{figure}[h]
    \centering
    \includegraphics[width=\textwidth]{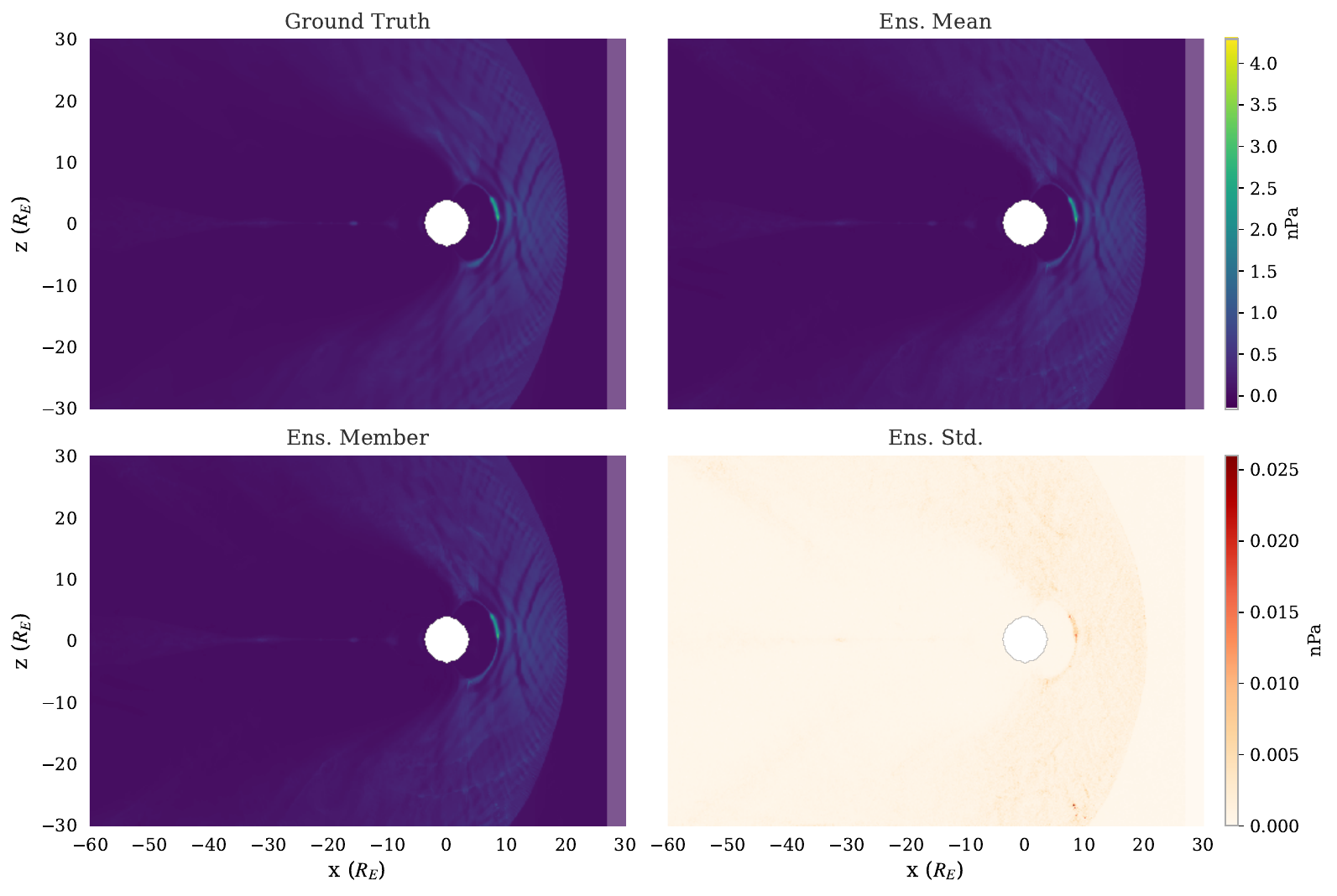}
    \caption{Plasma pressure $P$ at timestep 10 from Graph-EFM (hierarchical).}
    \label{fig:p}
\end{figure}

\begin{figure}[h]
    \centering
    \includegraphics[width=\textwidth]{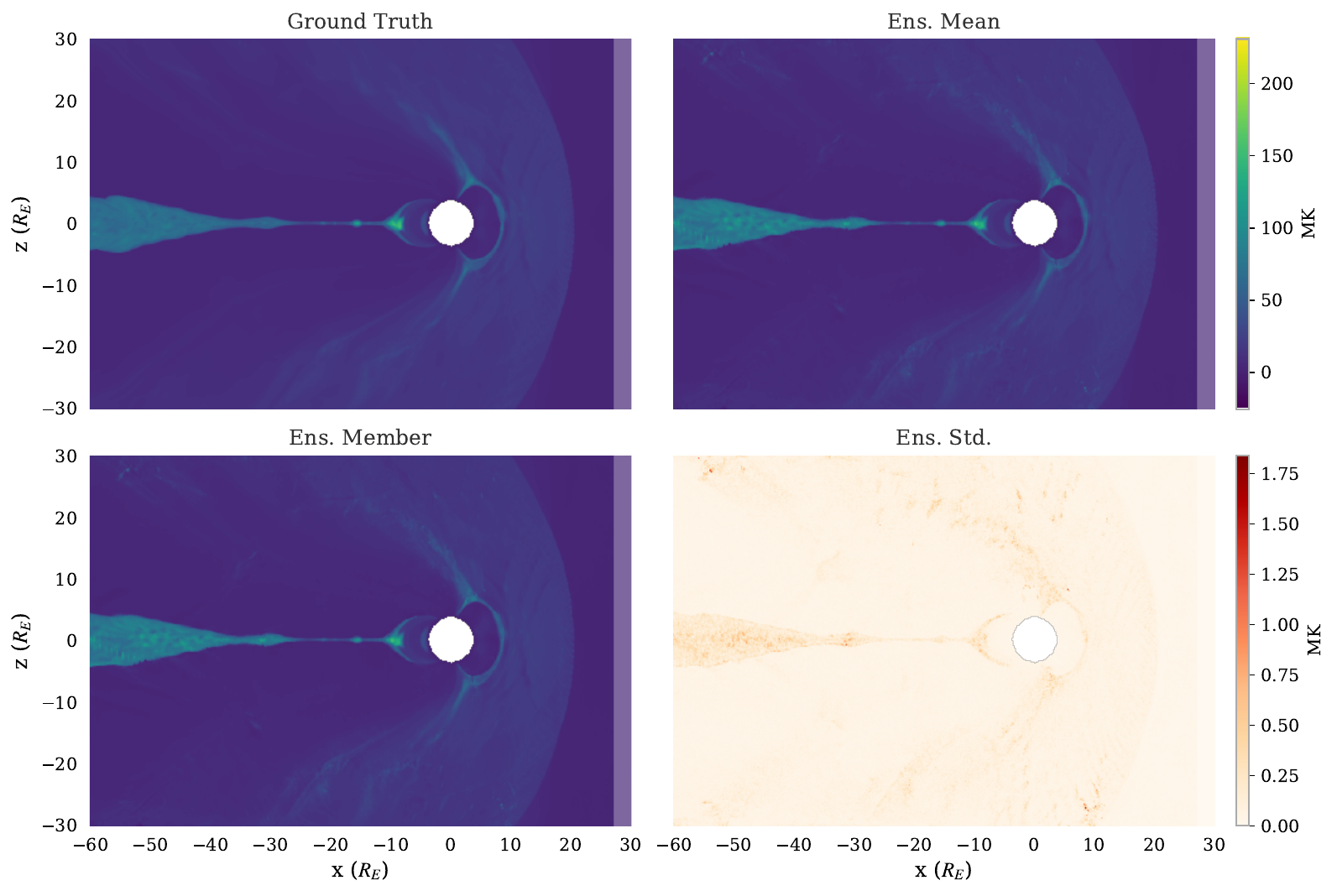}
    \caption{Plasma temperature $T$ at timestep 10 from Graph-EFM (hierarchical).}
    \label{fig:t}
\end{figure}

Visual inspection of an example test set forecast 10 steps ahead produced by the hierarchical Graph-EFM model in Fig.~\ref{fig:bx}--\ref{fig:t} reveals that it successfully captures complex and detailed magnetospheric dynamics. The ensemble standard deviation correctly localizes the highest uncertainty to physically active regions such as the bow shock, magnetotail, and the tail lobes. However, we also observe the emergence of spurious structures not present in the ground truth, for example in the predicted $B_y$ and $v_y$ components within the northern tail lobe. We attribute these artifacts primarily to the limited sample size of the training dataset.

To address these points, future work should focus on training with a larger dataset, which would make performance differences between models clearer and reduce artifacts. The ensemble calibration could be improved by increasing the CRPS weight during fine-tuning to enhance the spread. Finally, incorporating a rollout-based loss, i.e. optimizing over many future timesteps, can be a useful strategy to improve long-term stability of the emulators~\cite{lam2023graphcast}.

\clearpage

\bibliographyapp{references}
\bibliographystyleapp{unsrt}

\end{document}